\documentclass[prfluids, amsmath,amssymb,
aps]{revtex4-2}

\usepackage{graphicx}
\usepackage{epstopdf, epsfig}
\usepackage{bm}
\usepackage{amsmath}
\usepackage{amsfonts}
\usepackage{tikz-cd}

\usepackage{newtxtext}
\usepackage{newtxmath}
\usepackage{natbib}
\usepackage{hyperref}
\hypersetup{
    colorlinks = true,
    urlcolor   = blue,
    citecolor  = black,
}
\def\ba{\bm{a}}
\def\bx{\bm{x}}
\def\bu{\bm{u}}
\def\bk{\bm{k}}

\def\bphi{\bm{\varphi}}
\def\bXi{\bm{\Xi}}
\def\barbphi{\bar{\bar{\bphi}}}
\def\barphi{\bar{\bar{\varphi}}}
\def\barbu{\bar{\bar{\bm{u}}}}

\def\e{\mathrm{e}}
\def\d{\mathrm{d}}

\def\bnabla{\bm{\nabla}}
\def\bcdot{\bm{\cdot}}
\def\bXi{\bm{\Xi}}
\def\bchi{\tilde{\bm{\Xi}}}

\def\id{\mathrm{id}}
\def\i{\mathrm{i
}}

\def\bxi{\bm{\xi}}
\renewcommand{\phi}{\varphi}
\newcommand{\barL}[1]{\overline{#1}^\mathrm{L}}
\newcommand{\dt}[2]{\frac{\d#1}{\d#2}}

\newcommand{\RomanNumeralCaps}[1]
\linenumbers

\begin{document}
\preprint{APS/123-QED}

\title[Efficient Lagrangian averaging]{Efficient Lagrangian averaging with exponential filters}



\author{Abhijeet Minz}
 \author{Lois E. Baker}
\author{Jacques Vanneste}
\email[Email address for correspondence:]{J.Vanneste@ed.ac.uk}%

\affiliation{ 
School of Mathematics and Maxwell Institute for Mathematical Sciences, University of Edinburgh, Edinburgh EH9 3FD, UK
}%

\author{Hossein A. Kafiabad}
\affiliation{Department of Mathematical Sciences, Durham University, Durham DH1 3LE, UK
}%

\date{\today}

\begin{abstract}
Lagrangian averaging is a valuable tool for the analysis and modelling of multiscale processes in fluid dynamics. The numerical computation of Lagrangian (time) averages from simulation data is  challenging, however. It can be carried out by tracking a large number of particles or, following a recent approach, by solving a dedicated set of partial differential equations (PDEs). Both approaches are computationally demanding because they require an entirely new computation for each time at which the Lagrangian mean fields are desired.

We overcome this drawback by developing a PDE-based method that delivers Lagrangian mean fields for all times through the single solution of evolutionary PDEs.
This allows for an on-the-fly implementation, in which Lagrangian averages are computed along with the dynamical variables. This is made possible by the use of a special class of temporal filters whose kernels are sums of exponential functions. 

We focus on two specific kernels involving one and two exponential functions.
We implement these in the rotating shallow-water model and demonstrate their effectiveness at filtering out large-amplitude Poincar\'e waves while retaining the salient features of an underlying slowly evolving turbulent flow.  
\end{abstract}

\maketitle





\section{Introduction}

Many fluid dynamical phenomena involve temporal and spatial scales that cannot be fully resolved by measuring instruments or numerical simulations. Their modelling requires coarse graining, typically in the form of temporal or spatial averaging of the equations of motion, and the parameterisation of the impact of the unresolved scales. To this end, Lagrangian averaging, whereby averages are computed along fluid trajectories, has advantages over the more straightforward Eulerian averaging. 
These advantages stem from the preservation under Lagrangian averaging of the advective structure of the equations of motion, leading to valuable properties including the conservation of Lagrangian mean vorticity and circulation in the absence of dissipation and forcing \citep[e.g.][]{bret71,sowa72,andrews1978exact,grim84,holm02a,holm02b,buhler2014waves}.  Lagrangian averaging is also a valuable diagnostic tool for the decomposition of geophysical flows into waves and mean flow (see \citet[hereafter B24]{baker2024lagrangian} and references therein).

A conceptual framework for Lagrangian averaging is provided by the generalised Lagrangian mean (GLM) theory of \citet{andrews1978exact} \citep[see][for a comprehensive introduction]{buhler2014waves}. Its practical use, with temporal averaging substituted for the abstract averaging of GLM,   
has however been hampered by the challenges posed by the numerical computation of Lagrangian means from simulation data. Most implementations to date \citep{Nagai2015,Shakespeare2017c,Shakespeare2018,Shakespeare2019,Shakespeare2021a,bachmanParticleBasedLagrangianFiltering2020,jonesUsingLagrangianFiltering2023} rely on tracking a large number of particles. This is memory intensive, difficult to parallelise,
and often does not deliver Lagrangian mean fields in full agreement with the GLM definitions. Recently, 
\citet{kafiabad2022grid} proposed  a grid-based approach for the computation of the Lagrangian mean fields without tracking particles. 
Expanding on this \citet[hereafter KV23]{kafiabad2023computing} develop an approach based on the formulation of partial differential equations (PDEs) satisfied by the Lagrangian mean fields.  B24 extend the approach from the straightforward `top-hat' time averaging of KV23 to a broad class of linear filters obtained by convolution with an arbitrary kernel.
They moreover demonstrate the effectiveness of Lagrangian filtering for the  decomposition of flow variables into mean and perturbation.

The approach of KV23 and B24 remains costly in that independent evolutionary PDEs are integrated over (fast) time to derive the mean field at a single (slow) time. The process then needs to be repeated  for each slow time at which averaged fields are desired. There is, however, one class of filters that avoids this complication. 
The simplest member of this class is the exponential mean, with a truncated exponential as kernel. This enjoys the unique property of satisfying a closed (first order) differential equation. In the context of Lagrangian averaging, this leads to PDEs for the Lagrangian mean fields that share a single time variable with the dynamical equations and can therefore be solved in tandem with them, with modest overheads. (This benefit of the exponential mean was pointed out to the authors by O. B\"uhler.)  More generally, filters whose kernels are sums of exponentials lead to larger sets of PDEs, involving a number of auxiliary fields, that can similarly be solved alongside the dynamical equations.

In this paper,  we adapt the approach of KV23 to this class of sum-of-exponentials filters (described more fully in \S\ref{sec:exponentialmean}). We first focus on the (single) exponential mean (\S\ref{sec:computing}). Because of its remarkable simplicity, it offers a transparent introduction to PDE-based Lagrangian averaging (\S\ref{sec:singleexp}).  We demonstrate its practical value with an application to a rotating shallow-water flow that combines a slowly evolving turbulent flow with a large-amplitude fast Poincar\'e wave (\S\ref{sec:shallow}).  

The ability of the single exponential to filter out the fast wave is good, comparable to that of the top-hat filter of KV23, but small-amplitude fast oscillations remain in the mean fields that the more sophisticated low-pass filters of B24 would eliminate. We assess the capability of sum-of-exponential filters to  eliminate these by considering a special case, the 2nd-order Butterworth filter \citep[e.g.][]{tenoudji2016analog}, regarded as optimal among two-term exponential kernels (\S\ref{multi-filter}). We derive the necessary set of PDEs, implement them and find the resulting filtering to be highly effective for the rotating shallow-water example. Sum-of-exponentials kernels with more than two terms can in principle be implemented following a similar route. However, because the number of PDEs to be solved grows linearly with the number of terms, it may be impractical to go beyond 2 to 3 terms. Overall, the paper demonstrates the effectiveness of exponential filters which could make them the go-to method for the efficient computation of Lagrangian means.



\section{Lagrangian averaging and exponential filters} \label{sec:exponentialmean}

The GLM time average is defined as follows. Given the flow map $\bphi$, such that $\bphi(\ba,t)$ is the position at time $t$ of a fluid particle labelled by $\ba$, we first define a mean flow map by 
\begin{equation}
\barphi^i(\ba,t) = \overline{\phi^i(\ba,\cdot)}(t),
\label{eq:barphi}
\end{equation}
where the superscript $i$ indicates the coordinate and the overbar on the right-hand side denotes any moving time average evaluated at time $t$. We emphasise that this construction depends on choosing a coordinate representation of the flow map, since the average of the flow map itself (as opposed to its coordinate representation) is ill defined. We use the unconventional double-bar notation of \citet{gilbertGeometricApproaches2024} as a reminder of this. The Lagrangian mean of scalar field $g(\bx,t)$ is  defined by
\begin{equation}
\barL{g}(\barbphi(\ba,t),t) = \overline{g(\bphi(\ba,\cdot),\cdot)}(t).
\label{eq:barLg}
\end{equation}
In words, the Lagrangian mean $\barL{g}(\bx,t)$ is  the time average of $g$ following the trajectory of the particle whose mean position at $t$ is $\bx=\barbphi(\ba,t)$.

KV23 developed an approach for the computation of Lagrangian means of the form \eqref{eq:barLg} based on the solution of evolutionary PDEs. They implement their approach for the `top-hat' mean, that is, straightforward unweighted integration over a time interval $T$. This is extended by B24 to general linear filters, defined  for functions of time $h(t)$ as
\begin{equation}
\overline{h}(t) = \int_{-\infty}^\infty k(t-s) h(s) \, \d s
\label{eq:linfilt}
\end{equation}
for some kernel $k(t-s)$. The computational cost of these implementations is high because they require solving new initial-value PDE problems for each averaging time, that is, each time at which the Lagrangian mean is desired. In this paper, we examine a class of linear filters that dramatically reduce this cost. For this class, it is sufficient to solve a single system of PDEs to obtain the Lagrangian mean at all averaging times. 

The simplest filter in this class is the `exponential mean', with kernel
\begin{equation}
    k(t) = \alpha \e^{-\alpha t} \Theta(t),
\end{equation}
where $\Theta(t)$ is the Heaviside function and the parameter $\alpha>0$ is the inverse of an averaging time scale. The corresponding mean,
\begin{equation}
    \overline{h}(t) = \alpha \int_{-\infty}^t \e^{-\alpha(t-s)} h(s) \, \d s,
    \label{eq:expmean}
\end{equation}
has the benefit of being the solution to a first-order ODE, namely
\begin{equation}
    \dt{\overline{h}}{t}= \alpha (h - \overline{h}),
    \label{eq:expmeanODE}
\end{equation}
corresponding to a relaxation of $\bar h$ to $h$ with rate $\alpha$.
Eq.\ \eqref{eq:expmeanODE} makes it possible to compute the average on-the-fly as new values of $h(t)$ come in. For the computation of Lagrangian means following KV23, the exponential mean avoids the need to treat the averaging time as a (discretised) time variable distinct from the time variable used for the dynamical equations, as is required for the top-hat and other means. The exponential Lagrangian average is simply computed by solving additional PDEs at the same time steps as the dynamical equations.

The exponential mean has a broad frequency response instead of the sharp cutoff required to eliminate high frequencies while leaving low frequencies unaffected. There is therefore a need for more advanced filters which can also be implemented via the solution of differential equations. This is the case for linear filters of the form \eqref{eq:linfilt} with sum-of-exponentials kernels, that is, with
\begin{equation}
    k(t) = \sum_{i=1}^N a_i \e^{-\alpha_i t} \, \Theta(t),
    \label{eq:sumofexp}
\end{equation}
where  $\mathrm{Re} \, \alpha_i > 0$ and $a_i$ (such that $\sum_{i=1}^N a_i/\alpha_i = 1$).
The parameters $\alpha_i$ and $a_i$ can be selected to optimise properties of the filters such as frequency selectivity. For kernels of the form \eqref{eq:sumofexp}, the single first-order ODE \eqref{eq:expmeanODE} is replaced by a system of $N$ first-order ODEs for $\overline{h}(t)$ and $N-1$ auxiliary variables. In the context of Lagrangian averaging, this implies the solution of an extended set of PDEs. The computational cost, while not trivial, remains much lower than for the general linear filters of B24 if the Lagrangian average is required at high temporal resolution. 

In the remainder of the paper, we describe the Lagrangian implementation of the exponential mean \eqref{eq:expmean} and test its worth in shallow-water simulations. We examine the improvement brought about by a specific second-order filter ($N=2$ in \eqref{eq:sumofexp}) often regarded as optimal, namely the Butterworth filter \citep{tenoudji2016analog}. The generalisation to higher-order filters should be straightforward. 

\section{Exponential mean} \label{sec:computing}

\subsection{Formulation} \label{sec:singleexp}

Particularising \eqref{eq:barphi} and \eqref{eq:barLg} to the exponential mean 
\eqref{eq:expmean} gives the explicit forms
\begin{align}
    \barbphi(\ba,t) &= \alpha \int_{-\infty}^t \e^{-\alpha(t-s)} \bphi(\ba,s) \, \d s 
    \label{eq:barbphiexp} \\
    \textrm{and} \qquad \barL{g}(\barbphi(\ba,t),t) &= \alpha \int_{-\infty}^t \e^{-\alpha(t-s)} g(\bphi(\ba,s),s) \, \d s
    \label{eq:barLgexp}
\end{align}
for the mean map and Lagrangian mean of a scalar. (We abuse notation in \eqref{eq:barbphiexp} by using the same symbol $\barbphi$ for for both the mean map and the list of its components $\barphi^i$. We repeat this abuse with other maps in what follows.) The Lagrangian velocity $\barbu$ is defined naturally as the time derivative of the mean map:
\begin{equation}
    \partial_t \barbphi(\ba,t) = \barbu(\barbphi(\ba,t),t).
\end{equation}
Time differentiation of \eqref{eq:barbphiexp} gives
\begin{equation}
    \barbu(\bx,t) = \alpha ( \bXi(\bx,t) - \bx)
    \label{eq:barbuXi}
\end{equation}
on substituting $\bx=\barbphi(\ba,t)$. Here we introduce the so-called lifting map
\begin{equation}
    \bXi = \bphi \circ \barbphi^{-1}
    \label{eq:bXi}
\end{equation}
such that $\bXi(\bx,t)$ is the actual position at time $t$ of the fluid particle with mean position $\bx$. 
We note that $\barbu$ in \eqref{eq:barbuXi} also results from applying the exponential Lagrangian mean \eqref{eq:barLgexp} to each component of the velocity field $\bu$. This follows from the computation
\begin{align}
    &\alpha \int_{-\infty}^{t} \e^{-\alpha(t - s)} \bu(\bphi(\ba,s),s) \, \d s = \alpha \int_{-\infty}^{t} \e^{-\alpha(t - s)} \partial_s \bphi(\ba,s) \, \d s \nonumber \\
    & = \alpha \left[ \e^{-\alpha (t - s)} \bphi(\ba, s) \right]_{-\infty}^{t} - \alpha^2 \int_{-\infty}^t \e^{-\alpha (t - s)}  \bphi(\ba,s) \, \d s  \nonumber \\
    & = \alpha \left( \bphi(\ba, t) - \barbphi(\ba, t) \right)  
     = \alpha \left( \bXi(\bx,t) - \bx\right) \nonumber \\&= \barbu(\bx,t),
    \label{eq:barbu2}
\end{align}
where $\bx = \barbphi(\ba,t)$ and the last equality uses \eqref{eq:barbuXi}.

The lifting map $\bXi$ is required for computation of Lagrangian averages. It can be computed by solving the PDE obtained by differentiating the identity
\begin{equation}
    \bXi(\barbphi(\ba,t),t) = \bphi(\ba,t) 
\end{equation}
with respect to $t$ and substituting $\bx=\barbphi(\ba,t)$  to find
\begin{equation}
    \partial_t \bXi + \barbu \bcdot \bnabla \bXi = \bu \circ \bXi.
    \label{eq:XiPDE}
\end{equation}
Taking \eqref{eq:barbuXi} into account, this is a closed PDE for $\bXi$ that can be solved alongside the dynamical equations. 

Time differentiation of \eqref{eq:barLgexp} gives a PDE for the Lagrangian mean of the scalar field $g(\bx,t)$, namely
\begin{equation}
    \partial_t \barL{g} + \barbu \bcdot \bnabla \barL{g} = \alpha ( g \circ \bXi - \barL{g}). 
    \label{eq:barLgPDE}
\end{equation}
We solve this together with \eqref{eq:XiPDE} since $\bXi$ is required both for $\barbu$ on the left-hand side and for the composition on the right-hand side. As initial conditions we use $\bXi(\bx,0)=\bx$ and $\barL{g}(\bx,0) = 0$. This leads to $\barL{g}(\bx,t)$, at least for $t$ large enough that the integration limit $t \to - \infty$ in \eqref{eq:barLgexp} can be replaced by $t=0$. After discretisation, the computation of the right-hand sides of  \eqref{eq:XiPDE} and \eqref{eq:barLgPDE} requires interpolations to evaluate $\bu \circ \bXi$ and $g \circ \bXi$.

Eqs.\ \eqref{eq:barbuXi}, \eqref{eq:XiPDE} and \eqref{eq:barLgPDE} are key results of this paper. They provide a route for an efficient, easy-to-implement computation of Lagrangian averages leveraging the unique properties of the exponential mean. Eq.\ \eqref{eq:barLgPDE} in particular can be thought of as a Lagrangian version of the ODE \eqref{eq:expmeanODE} for the exponential mean. 

\subsection{Rotating shallow-water example} \label{sec:shallow}


We compute Lagrangian means in a simulation of a turbulent flow interacting with a Poincaré wave in a rotating shallow-water model \citep[e.g.][]{vallis2017atmospheric,zeitlin2018geophysical}. We use the non-dimensional equations for rotating shallow water  with characteristic length $L$, characteristic velocity $U$ and characteristic time $T = L/U$, that is, 
\begin{subequations}  \label{eq:shallow water}
\begin{align}
           \partial_t{\bu} + \bu \bcdot \bnabla \bu + Ro^{-1} \hat{\bm{z}} \times \bu &= - Fr^{-2} \bnabla h, \label{eq:shallow water (a)} \\
        \partial_t{h} + \bnabla \bcdot (h \bu) &= 0, 
\end{align}
\end{subequations}
which introduces the Rossby and Froude numbers
\begin{equation}
    Ro = \frac{U}{fL} \quad \textrm{and} \quad Fr =\frac{U}{\sqrt{gH}}. 
\end{equation}
Here $g$ is the gravitational acceleration, $f$ the Coriolis parameter, $H$ the mean depth and $\hat{\bm{z}}$ the vertical unit vector. 

We solve the dynamical equations \eqref{eq:shallow water} together with the Lagrangian mean equations \eqref{eq:XiPDE} and \eqref{eq:barLgPDE} using a pseudospectral discretisation. To solve exclusively for doubly periodic fields, we express the lifting map $\bXi$ in terms of the (doubly periodic) displacement map $\bxi(\bx, t) = \bXi(\bx, t) - \bx$ replacing \eqref{eq:XiPDE} by
\begin{equation}
    \partial_t \bxi + \barbu \bcdot \bnabla \bxi = \bu \circ (\id + \bxi) - \barbu, \label{eq:PDE_smallxi}
\end{equation}
where $\id$ denotes the identity map and $\bxi(\bx,0)=0$, and \eqref{eq:barLgPDE} by 
\begin{equation}
    \partial_t \barL{g} + \barbu \bcdot \bnabla \barL{g} = \alpha (g \circ (\id + \bxi) - \barL{g}) \label{eq:LM_vort}.
\end{equation}
The Lagrangian mean velocity is deduced from $\bxi$ as $\barbu = \alpha \bxi$ obtained from \eqref{eq:barbuXi}.

We use a modified version of the code developed by KV23. We discretise \eqref{eq:shallow water}, \eqref{eq:PDE_smallxi} and \eqref{eq:LM_vort} using $256^2$ grid points/Fourier modes, use an RK4 integrator with time step $\Delta t =  0.005$ and integrate over the time range $0 \le t \le T = 50$. For numerical stability, we employ 2/3 dealiasing and add hyperviscous dissipation to the momentum shallow-water equation \eqref{eq:shallow water (a)} which amounts to multiplying the Fourier transform $\hat{\bu}(\bk,t)$ of the velocity by $\exp(-\kappa |\bk|^8 \Delta t)$ at each time step. We take $\kappa = 2.6 \times 10^{-14}$. 
We achieve a stable numerical solution of the Lagrangian mean equations \eqref{eq:PDE_smallxi} and \eqref{eq:LM_vort} without any dissipation. However, a form of dissipation may be required to ensure numerical instability in other flow configurations.
We use bilinear interpolation to evaluate $\bu$ and $g$ at the position $\bx + \bxi(\bx, t)$ as required for  \eqref{eq:PDE_smallxi} and \eqref{eq:LM_vort}.



We initialise the model with the same initial condition as in KV23 superimposing a geostrophic turbulent flow (obtained by prior solution of an incompressible two-dimensional Navier–Stokes model for the velocities and initialisation of the height to be in geostrophic balance) and a rightward travelling mode-$1$ Poincar\'e wave. 
We take the root-mean square velocity of the geostrophic flow as characteristic velocity $U$ and the length scale of the first Fourier mode as characteristic length $L$ so that the doubly periodic domain is $[0 ,2\pi]^2$. The Rossby and Froude number of the geostrophic flow are $Ro = 0.1$ and $Fr = 0.5$. With these parameters, the frequency $\omega = (Ro^{-2} + Fr^{-2})^{\frac{1}{2}}$ of the mode-$1$ Poincaré wave is $\omega = 10.2$ corresponding to a period of $0.62$ time units.
The wave fields in the absence of flow are given by
\begin{equation}
    u' = a\ \cos(x - \omega t), \quad v' = \frac{a}{\omega Ro} \ \sin(x - \omega t), \quad h' = \frac{a}{\omega} \ \cos(x - \omega t). \label{eq:wave_amp}
\end{equation}
At the initial time $t=0$, we add these to the geostrophic flow field. We take the amplitude $a=-1$ so that the maximum wave velocity is as large as the geostrophic flow root-mean-square velocity.

\begin{figure}
\begin{center}
   \includegraphics[width = 1.00\textwidth]{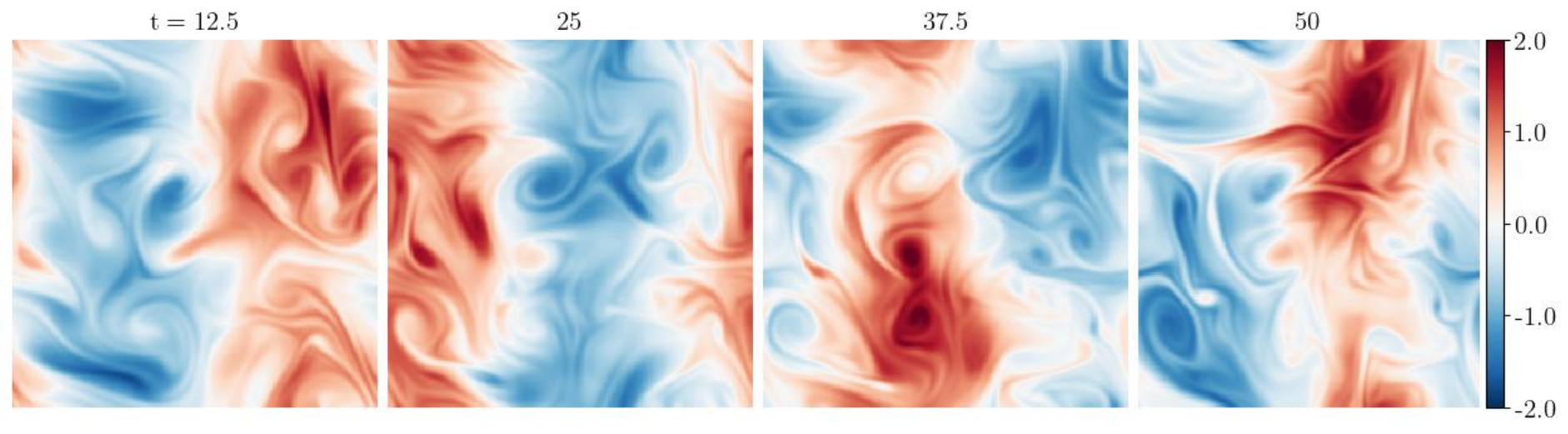} \\
   \includegraphics[width = 1.00\textwidth]{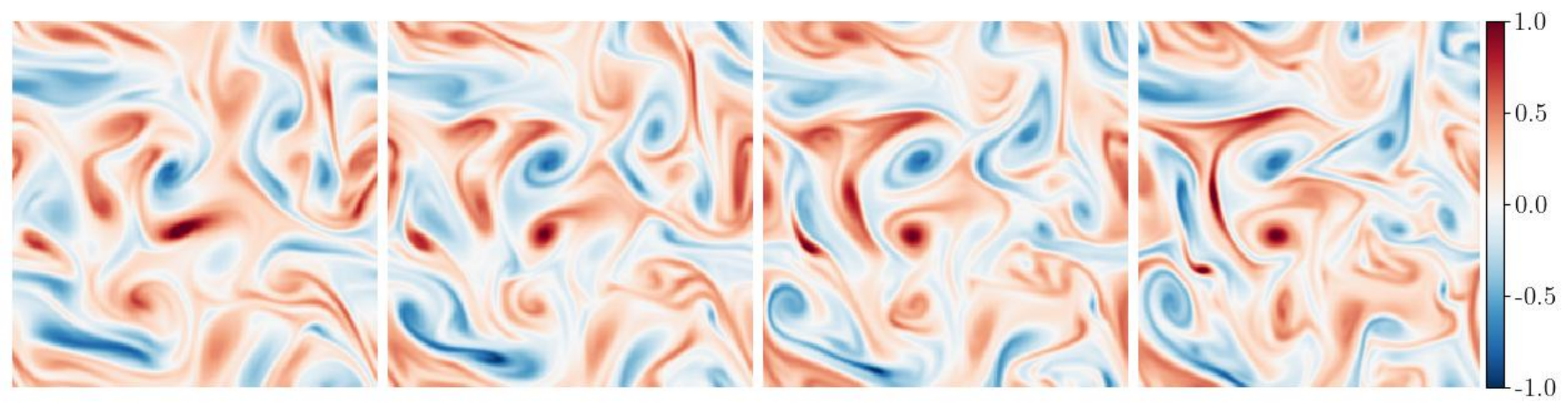}\\
   \includegraphics[width = 1.00\textwidth,]{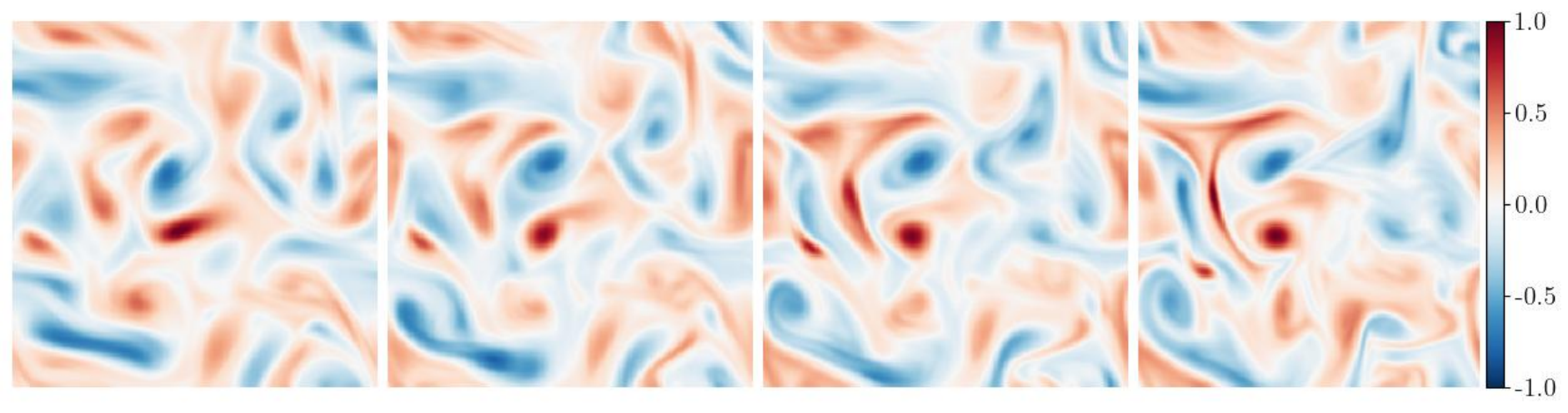}
  \caption{Vorticity field in the rotating shallow-water simulation at $t=12.5, \, 25,\, 37.5$ and $50$: instantaneous vorticity $\zeta$ (top row), Lagrangian mean vorticity $\barL{\zeta}$ (middle row) and  Eulerian mean vorticity $\bar{\zeta}$ (bottom row). The exponential mean with inverse averaging time scale $\alpha = 0.5$ is used for the mean vorticity fields in the middle and bottom rows. See also the animation \href{http://www.maths.ed.ac.uk/~vanneste/exponentialMean/movie1.mp4}{movie 1}.}
 \label{fig: vortex_sim_1}
 \end{center}
\end{figure}

We first consider the Lagrangian mean of the relative vorticity $\zeta = \partial_x v - \partial_y u$ and thus set $g = \zeta$ in \eqref{eq:LM_vort}.
The top row of figure \ref{fig: vortex_sim_1} shows $\zeta$ at $t=12.5, \, 25, \, 37.5$ and $50$. The mode-1 wave which we aim to filter out is a dominant feature. It is distorted by the flow consisting of vortices and filaments familiar in two-dimensional (quasi-geostrophic) turbulence. The middle and bottom rows compare the Lagrangian and Eulerian means $\barL{\zeta}$ and $\bar{\zeta}$ computed at each $t$ with an inverse averaging time scale $\alpha = 0.5$. Both means filter out the wave satisfactorily, but the Eulerian mean blurs the small-scale vorticity structures of the flow as a result of the rapid advection by the velocity field associated with the wave. The Lagrangian mean eliminates this advection by construction.  

\begin{figure}
\begin{center}
  \includegraphics[width = 1\textwidth]{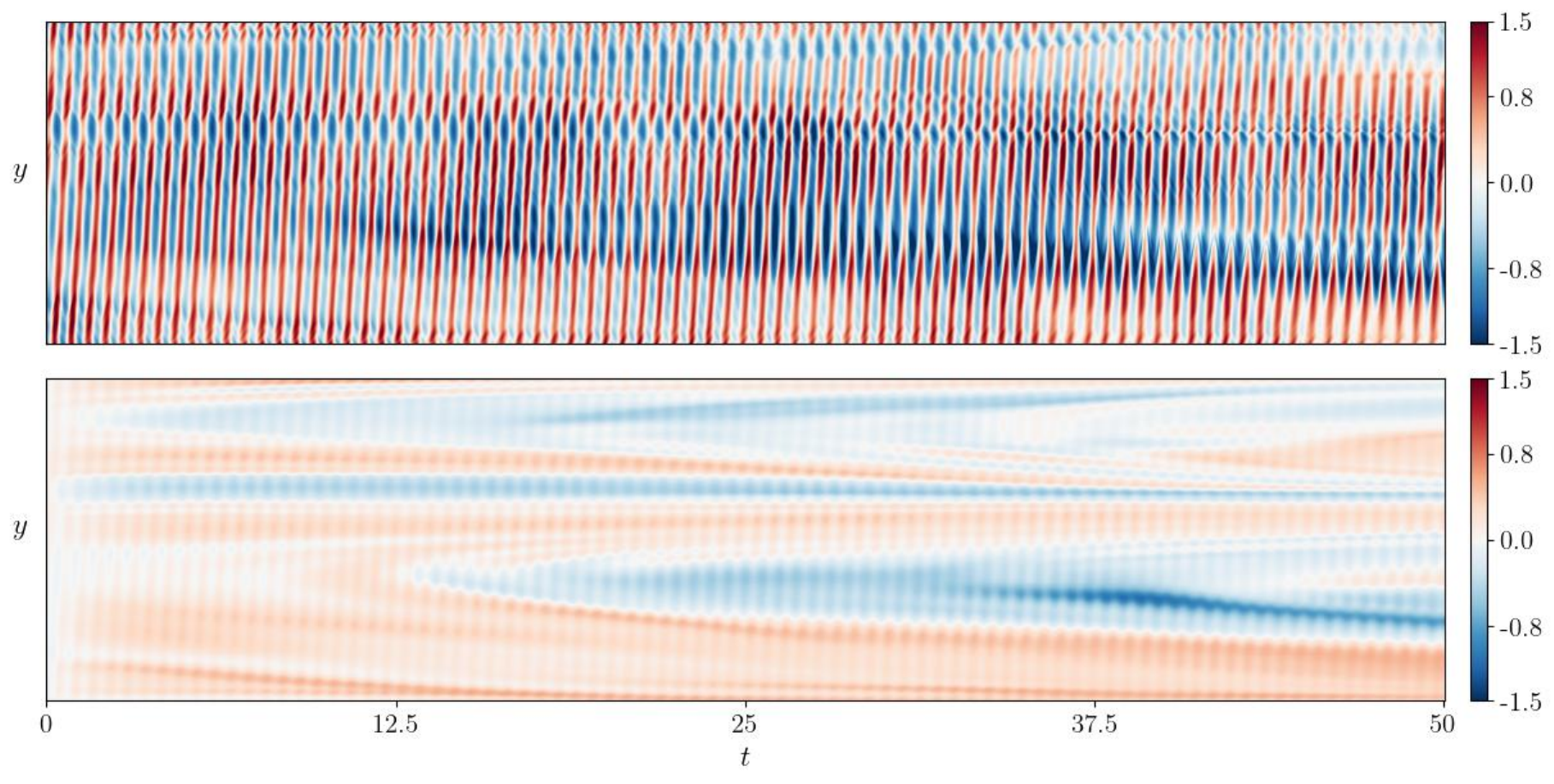}
  \caption{Vorticity $\zeta$ (top)  and its exponential Lagrangian mean $\barL{\zeta}$ with $\alpha=0.5$ (bottom) as functions of $t$ and $y$ for $x=0.24$ in the simulation in figure \ref{fig: vortex_sim_1}.}
 \label{fig: vortex_sim_HM}
 \end{center}
\end{figure}

A qualitatively similar Lagrangian mean field is obtained using the top-hat mean of KV23. However, the exponential mean provides a much simpler way of computing a Lagrangian mean over an entire time interval rather than a single snapshot. In particular, since the exponential Lagrangian mean computation shares the same (small) time step as the solution of the dynamical equations, it resolves the fast wave time scales in the system, making it straightforward to examine the wave signal that remains after averaging. 

We illustrate this with figure \ref{fig: vortex_sim_HM}, which shows a Hovm\"oller diagram of $\zeta$ and $\barL{\zeta}$ at fixed $x=0.24$, and with an animation of the evolution of $\zeta$, $\barL{\zeta}$ and $\bar{\zeta}$ (\href{http://www.maths.ed.ac.uk/~vanneste/exponentialMean/movie1.mp4}{movie 1}).
The wave signal in the instantaneous field $\zeta$ is strongly reduced by the exponential averaging, yielding a mean field $\barL{\zeta}$ that captures the slow dynamics well, but the wave signal is not eliminated completely. This is not surprising: the exponential mean is a low-pass filter with a broad frequency response which reduces rather than eliminates high frequencies  (its Fourier-domain transfer function at frequency $\Omega$ is $(\alpha - \i \Omega)^{-1}$). This limitation of the exponential mean is also evident in the relation $\barbu = \alpha \bm{\xi}$ between the (slow) mean velocity $\barbu$ and the (fast) displacement $\bm{\xi}$. 
The notional slowness of $\barbu$ stems only from the property $|\barbu| \ll |\partial_t \bm{\xi}|$, which is obeyed provided that $\alpha/\omega \ll 1$. More advanced filters can meet stricter slowness conditions such as  $|\partial^n_t \barbu| \sim (\alpha/\omega)^n |\partial^{n+1}_t \bm{\xi}|$ for  $n=0,\, 1,\, 2,\cdots$.
\begin{figure*}
\includegraphics[width = 1\textwidth]{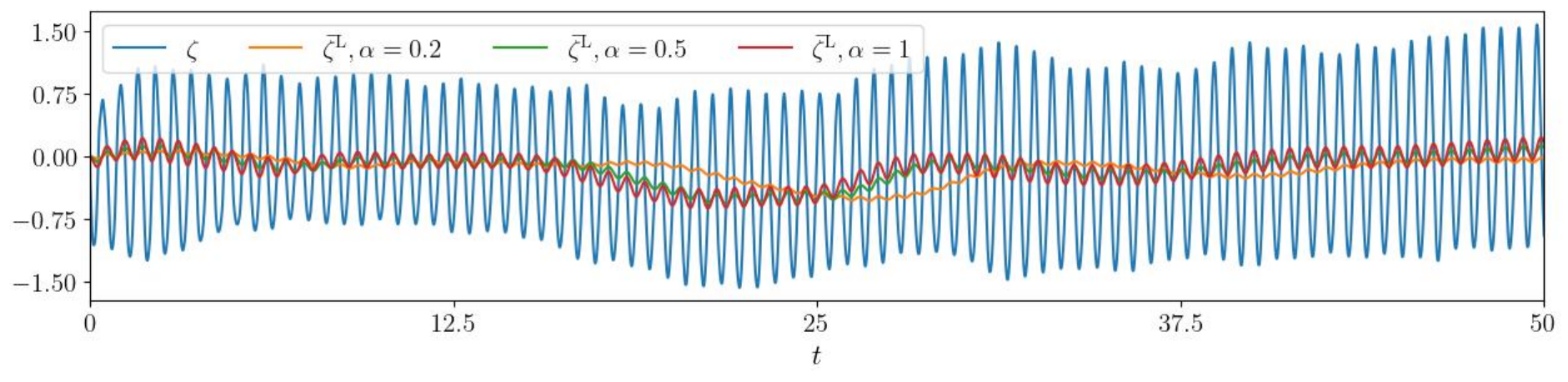}
\caption{Vorticity at $(x,y)= (1.2,1.2)$ as a function of $t$: the instantaneous value $\zeta$ is compared with the exponential Lagrangian average with $\alpha=0.2, \, 0.5$ and $1$.}
 \label{fig:zetafixedpoint}
\end{figure*}



The choice of $\alpha$ determines which range of frequencies are filtered out effectively by the exponential mean. Figure \ref{fig:zetafixedpoint} illustrates the effect of varying $\alpha$ by showing $\zeta$ and $\barL{\zeta}$ at the fixed position $(x,y)=(1.2,1.2)$ as a function of time for three values of $\alpha$, all for which such that $\alpha/\omega \ll 1$. 
Clearly, decreasing $\alpha$ reduces the magnitude of the fast oscillations induced by the wave. For the smallest value of $\alpha$, $\alpha=0.2$, $\barL{\zeta}$ differs markedly from the (Eulerian) average that is estimated by eye from the instantaneous $\zeta$; this is a reminder that Lagrangian averaging is non-local in space and depends on the field at positions other than where it is evaluated. In applications, the choice of $\alpha$ should be guided by the use made of the Lagrangian averaging. In particular, when the aim is to assess the impact of waves on the mean flow, there is a trade off between the requirement to filter out fast waves and the need to retain the dynamically significant time scales of the flow. 
Our experimentations show that, for the flow considered, $\alpha=0.5$ is a good compromise. This suggests that, more generally, exponential averaging with inverse averaging time $\alpha$ is effective in filtering out waves of frequencies $\omega \gtrsim 20 \alpha$.

\begin{figure}
\begin{center}
 \includegraphics[width = 1.00\textwidth]{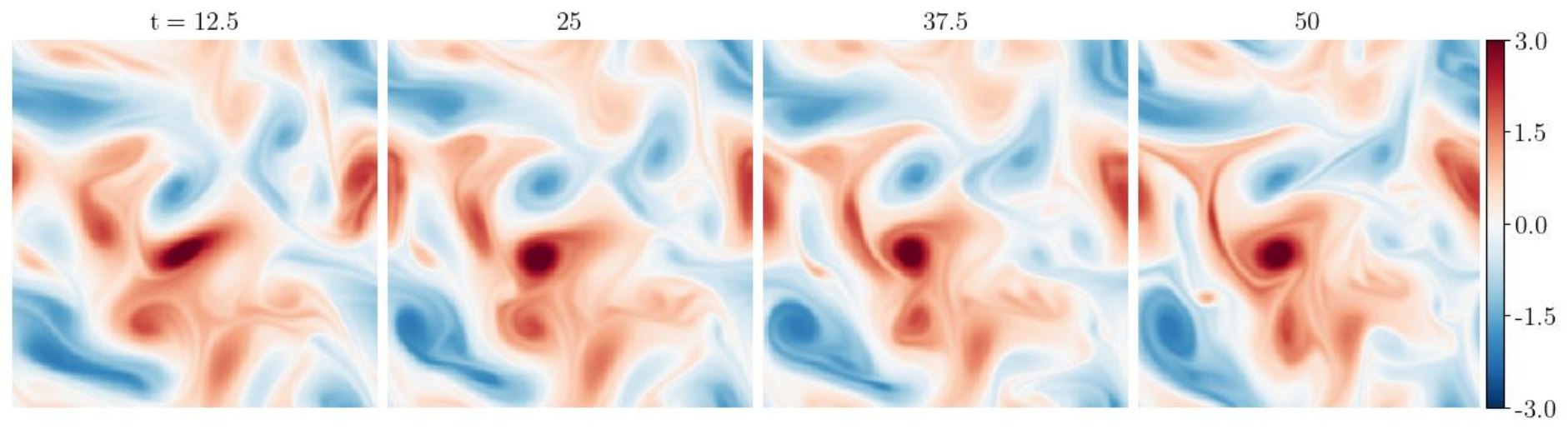} \\
 \includegraphics[width = 1.00\textwidth]{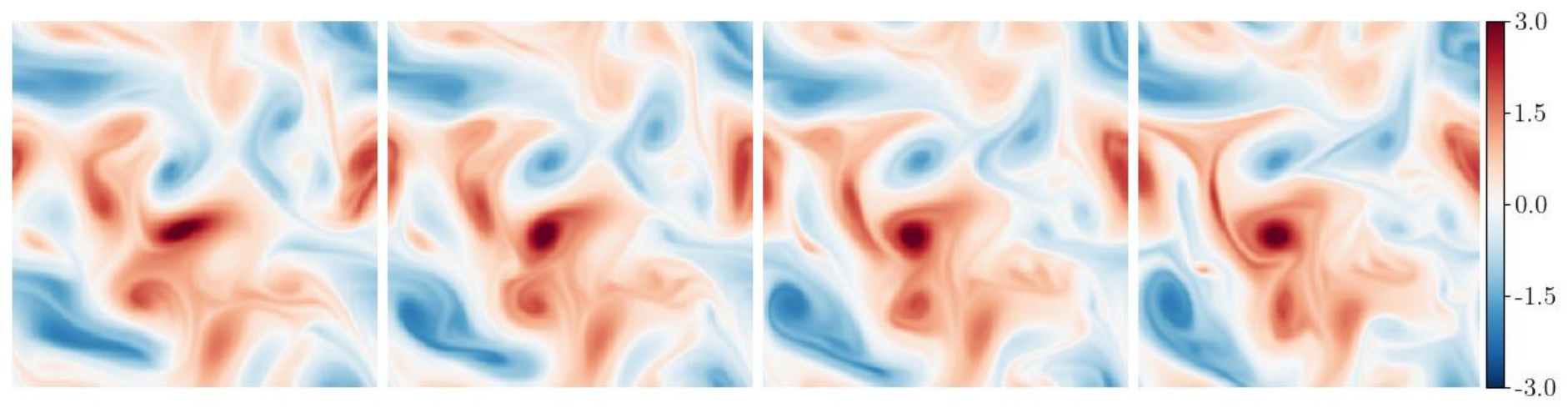}  \\
 \includegraphics[width = 1.00\textwidth]{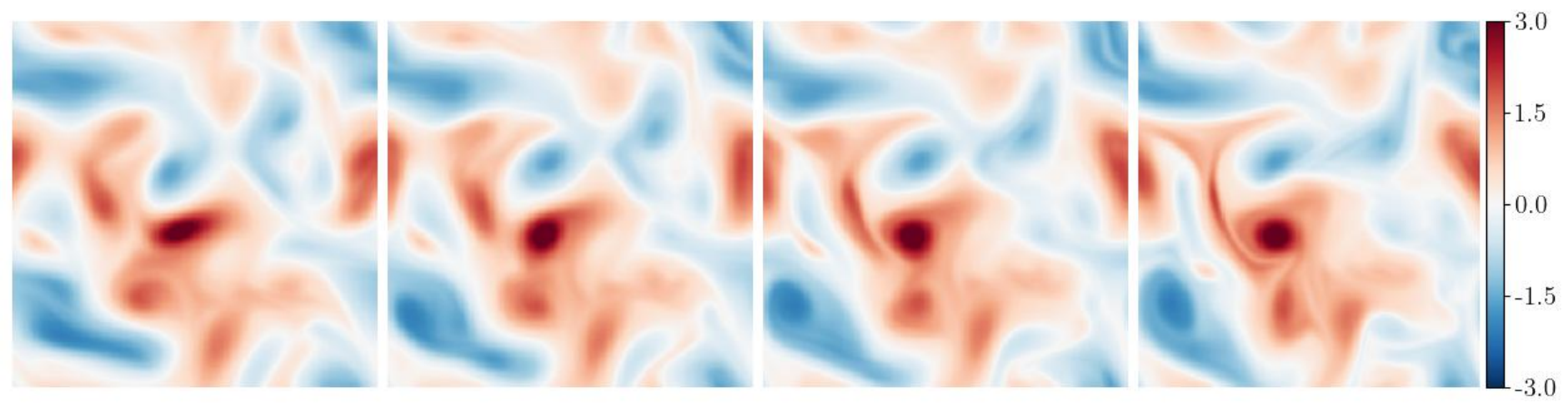}  
 
 \caption{Same as figure \ref{fig: vortex_sim_1} but for the potential vorticity anomaly $q$ at $t=12.5,\, 25, \, 37.5$ and $50$: $q$  (top row),  $\barL{q}$ (middle row) and  $\bar q$ (bottom row). See also the animation \href{http://www.maths.ed.ac.uk/~vanneste/exponentialMean/movie2.mp4}{movie 2}.}
 \label{fig:PV}
 \end{center}
\end{figure}

 A scalar of particular interest is the potential vorticity (PV)
\begin{equation}
    q = \frac{Ro^{-1} + \zeta}{h} - Ro^{-1}
\end{equation}
which we give here in non-dimensional form with the constant background PV subtracted. Because the PV of the Poincar\'e wave is zero (at a linear level), it is impacted by the wave in a very different way to the relative vorticity $\zeta$: advection by the velocity field associated with the wave induces a fast oscillation of PV structures that are otherwise unaffected. These oscillations limit the usefulness of the Eulerian mean PV compared with the Lagrangian mean PV. We illustrate this in figure \ref{fig:PV}, which shows snapshots of the PV $q$ and of its exponential Lagrangian mean $\barL q$ and of the Eulerian mean $\bar q$. Their dynamics are  more evident in the animations available in \href{http://www.maths.ed.ac.uk/~vanneste/exponentialMean/movie2.mp4}{movie 2}. Because the instantaneous PV is rapidly advected back and forth by the wave,
its Eulerian mean is blurred. In contrast, the Lagrangian mean retains all the spatial features of the instantaneous field while filtering out the fast advective oscillations. In the absence of dissipation, the material conservation of $q$ implies that $q = q_0 \circ \bphi^{-1}$, with $q_0$ the initial PV field, hence $\barL{q} = \overline{q \circ \bXi}$ is the average of $q_0 \circ \barbphi^{-1}$ and approximately equal to $q_0 \circ \barbphi^{-1}$ itself insofar as $\barbphi^{-1}$ can be regarded as slow. Therefore, the Lagrangian mean PV is approximately a rearrangement of the instantaneous PV, thus sharing the same extrema and topology.

\section{Sum-of-exponential filters: 2nd-order Butterworth mean} \label{multi-filter}

\subsection{Formulation} \label{sec:multi-formulation}

We now demonstrate how the residual wave signal observed in the mean fields in \S\ref{sec:exponentialmean} can be dramatically reduced by replacing the exponential filter by an improved filter whose kernel takes the sum-of-exponentials form \eqref{eq:sumofexp}.
When applied to functions of time, sum-of-exponentials filters can be computed via the solution of a system of first order ODEs extending the  single first-order ODE \eqref{eq:expmeanODE} of the exponential mean. In the context of Lagrangian averaging, they provide Lagrangian mean fields via the solution of a system of PDEs extending \eqref{eq:XiPDE}--\eqref{eq:barLgPDE}.
We illustrate this with a specific two-term kernel corresponding to the
2nd-order Butterworth filter \citep{tenoudji2016analog}.  Butterworth filters are optimal in that, for a given number of terms, they have a maximally flat frequency response at $0$ frequency. 

The 2nd-order Butterworth filter is usually characterised by the Laplace transform $K(s)$ of its kernel $k(t)$, given by
\begin{equation}
    K (s) = \frac{\alpha^2}{s^2 + \sqrt{2} \alpha s + \alpha^2},\label{eq:BWTransfunc}
\end{equation}
so that
\begin{equation}
    k(t) = \sqrt{2} \alpha  \,  \e^{-\alpha t/\sqrt{2}} \sin(\alpha t / \sqrt{2}) \, \Theta(t). 
    \label{eq:Bk}
\end{equation}
This corresponds to the sum-of-exponential kernel \eqref{eq:sumofexp} with $N=2$, $\alpha_1 = \alpha_2^* = (1+\i)\alpha /\sqrt{2}$ and $a_1 = - a_2 = \i \alpha/ \sqrt{2}$.  The only parameter, $\alpha$, is an inverse averaging time, as in the case of the exponential filter.

This filter can be implemented via the solutions of two ODEs as follows. For a function $h(t)$ of time alone, we write the ODEs as
\begin{equation}\label{eq:multi-filter0}
        \frac{d}{dt}\begin{bmatrix} \Tilde{h}(t)  \\  \bar{h}(t) \end{bmatrix} = -\alpha \ A \begin{bmatrix} \Tilde{h}(t)  \\  \bar{h}(t)
    \end{bmatrix} + \alpha \begin{bmatrix} h(t) \\ 0 \end{bmatrix},
\end{equation}
where $\tilde{h}(t)$ is an auxiliary variable and $A$ a positive definite matrix to be determined. We insist that when $h(t)=1$, $\bar h(t)=\tilde h(t) = 1$, so that both $\bar h(t)$ and $\tilde h(t)$ can be interpreted as mean variables. This leads to the first constraint
\begin{equation}
    A \begin{bmatrix} 1 \\ 1 \end{bmatrix} = \begin{bmatrix} 1 \\ 0 \end{bmatrix}.
    \label{eq:c1}
\end{equation}
A second constraint follows from the requirement that $\bar h(t)$ is the convolution of $h(t)$ with the kernel \eqref{eq:Bk} or, in terms of Laplace transforms, that
\begin{equation}
    \bar H(s) = K(s) H(s),
\end{equation}
with $H(s)$ and $\bar H(s)$  the Laplace transforms of $h(t)$ and $\bar h(t)$.
Taking the Laplace transform of \eqref{eq:multi-filter0}, we write this explicitly as 
\begin{equation}
[(A + s I / \alpha)^{-1}]_{21} = K(s), 
\label{eq:c2}
\end{equation}
where $I$ is the identity matrix and the subscript 21 denotes the entry in the second row and first column. A short computation shows that the matrix $A$ satisfying \eqref{eq:c1}--\eqref{eq:c2} is
\begin{equation}\label{eq:multi_Matrix_A}
 A = \begin{bmatrix} \sqrt{2} - 1 & 2 - \sqrt{2} \\ -1 & 1 \end{bmatrix}.
\end{equation}

We obtain PDEs for the computation of Lagrangian means using the Butterworth filter by adapting the construction above to the flow map $\bphi(\ba, t)$ and an arbitrary scalar field $g(\bx,t)$. For the flow map, we write
\begin{equation}
    \partial_t \begin{bmatrix} \tilde\bphi(\ba,t) \\ \barbphi(\ba,t) \end{bmatrix} = - \alpha A \begin{bmatrix} \tilde\bphi(\ba,t) \\ \barbphi(\ba,t) \end{bmatrix} + \alpha \begin{bmatrix} \bphi(\ba,t) \\ 0 \end{bmatrix}, 
    \label{eq:phiA}
\end{equation}
where $\barbphi$ is the mean map and $\tilde{\bphi}$ is an auxiliary map. Defining the mean and auxiliary vector fields 
\begin{equation}
    \barbu = \partial_t \barbphi  \circ \barbphi^{-1} \quad \textrm{and} \quad \tilde{\bu}  =  \partial_t \tilde{\bphi} \circ \barbphi^{-1}
\end{equation}
we rewrite \eqref{eq:phiA} as
\begin{equation}
    \begin{bmatrix} \tilde{\bu} (\bx,t) \\ \barbu(\bx,t)  \end{bmatrix} = - \alpha A \begin{bmatrix} \bchi(\bx,t) \\ \bx \end{bmatrix} + \alpha \begin{bmatrix} \bXi(\bx,t) \\ 0 \end{bmatrix},\label{eq:multi-velocity}
\end{equation}
where 
\begin{equation}
   \bchi = \tilde{\bphi} \circ \barbphi^{-1} \quad \textrm{and} \quad  \bXi = \bphi \circ \barbphi^{-1}
   \label{eq:chi}
\end{equation} 
are perturbation maps.
Differentiating \eqref{eq:chi} with respect to $t$ yields
\begin{equation}
    \partial_t \bchi + \barbu \bcdot \bm\nabla \bchi = \tilde{\bu} 
   \quad \textrm{and} \quad \partial_t \bXi + \barbu \bm\cdot \bm\nabla \bXi = \bu \circ \bXi.\label{eq:Multi-Xi_PDE_2}
\end{equation}
Eqs.\ \eqref{eq:multi-velocity} and \eqref{eq:Multi-Xi_PDE_2} form a closed system for the perturbation maps $\bchi$ and $\bXi$.

The Lagrangian mean $\barL{g}(\bx,t)$ of a scalar field $g(\bx,t)$ is deduced from the system
\begin{equation}\label{eq:multi-scalr_avg}
    \partial_t \begin{bmatrix} \tilde{g}(\barbphi(\ba,t),t) \\  \barL{g}(\barbphi(\ba,t),t) 
     \end{bmatrix} = -\alpha \ A \begin{bmatrix} \tilde{g} (\barbphi(\ba,t),t) \\ \barL{g}(\barbphi(\ba,t),t)
    \end{bmatrix} + \alpha \begin{bmatrix} g(\bphi(\ba,t),t) \\ 0 \end{bmatrix},
\end{equation}
where $\tilde{g}$ is an auxiliary field and the label $\ba$ is fixed.  Now, \eqref{eq:multi-scalr_avg} can be rewritten in terms of PDEs, incorporating the definition of matrix $A$ from \eqref{eq:multi_Matrix_A}, as follows:
\begin{subequations} \label{eq:multi-scalar_PDE} 
\begin{align}
    \partial_t \tilde{g}  + \barbu \bm\cdot \bm\nabla \tilde{g}  &= -\alpha \left((\sqrt{2} -1)\tilde{g} + (2 - \sqrt{2})\barL{g} + g \circ \bXi \right) \label{eq:multi-scalar_PDE1}\\
    \partial_t \barL{g} + \barbu \bm\cdot \bm\nabla \barL{g} &= -\alpha \left(-\tilde{g} + \barL{g} \right)\label{eq:multi-scalar_PDE2} 
\end{align}
\end{subequations}
We solve this alongside  \eqref{eq:Multi-Xi_PDE_2} as both $\bchi$ and $\bXi$ are needed to determine mean velocity field $\barbu$ in \eqref{eq:multi-velocity}. Moreover, $\bXi$ is crucial for evaluating right-hand side, specifically $g \circ \bXi$, in \eqref{eq:multi-scalar_PDE1}. The initail condition is $\bXi(\bx, 0) = \bx$, $\bchi(\bx, 0) = \bx$,  $\tilde{g}(\bx,0) = 0$ and $\barL{g}(\bx,0) = 0$ to obtain $\barL{g}(\bx,t)$ which, after an adjustment period, can be interpreted as the 2nd-order Butterworth Lagrangian mean. After discretisation, the computation requires interpolations to evaluate $\bu \circ \bXi$ and $g \circ \bXi$ at each time step. The number of interpolations is the same as for the (single) exponential mean.

\subsection{Rotating shallow-water example}  \label{sec{multi-example}}

\begin{figure}
\begin{center}
  \includegraphics[width = 1\textwidth]{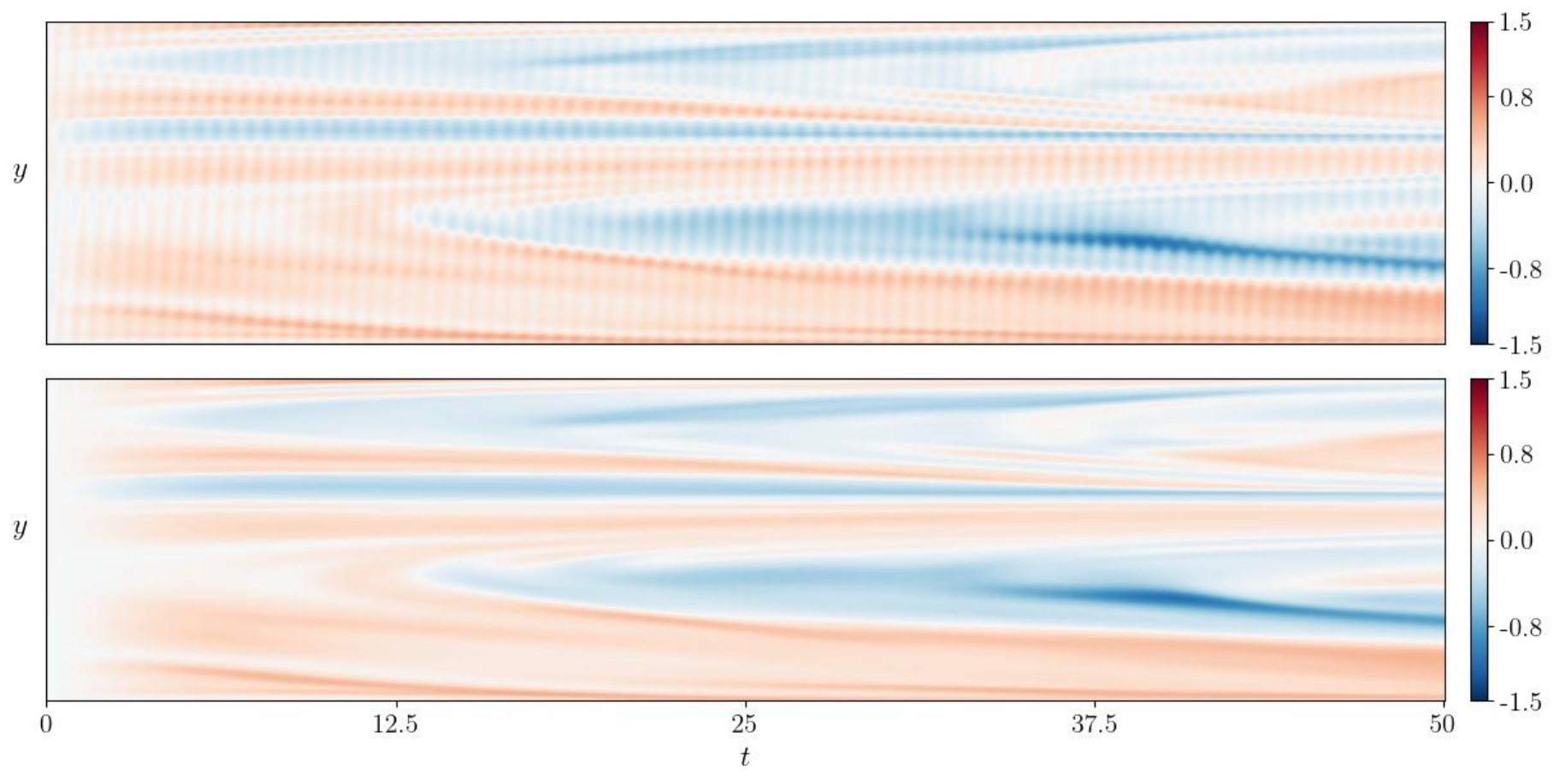}
  \caption{Lagrangian mean vorticity $\barL \zeta$ obtained with the exponential mean (top, reproducing the bottom panel of figure \ref{fig: vortex_sim_1}) and the 2nd-order Butterworth filter (bottom) as functions of $t$ and $y$ for $x=0.24$}
 \label{fig: multi-vortex_sim_HM}
 \end{center}
\end{figure}

\begin{figure}
\includegraphics[width = 1\textwidth]{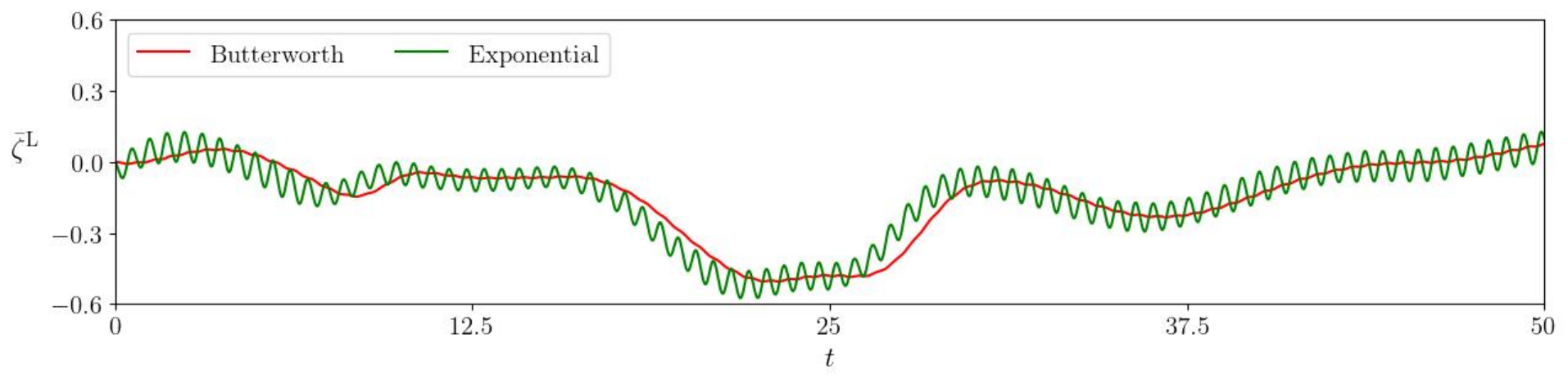}
\caption{Lagrangian mean vorticity $\barL \zeta$ obtained with the (single) exponential mean and the 2nd-order Butterworth filter as functions of $t$ for $(x,y)= (1.2,1.2)$. }
 \label{fig:multi-zetafixedpoint}
\end{figure}

We compute Lagrangian means using the Butterworth filter for the rotating shallow water  in \S\ref{sec:shallow}. To implement \eqref{eq:multi-velocity} and \eqref{eq:multi-scalar_PDE}  in a periodic domain, we express $\bXi$ and $\bchi$ in terms of periodic displacement maps, defined as $\bxi(\bx,t) = \bXi(\bx,t) - \bx$ and $\tilde{\bxi}(\bx,t) = \bchi(\bx,t) - \bx$. Eqs.\  \eqref{eq:multi-velocity} and \eqref{eq:Multi-Xi_PDE_2} then become
\begin{subequations} \label{eq:multi_xi}
\begin{align}
\tilde{\bu} = \alpha (\bxi - & (\sqrt{2} -1)  \tilde\bxi), \quad 
\barbu = \alpha \tilde\bxi, \\
    \partial_t \tilde\bxi + \barbu \bcdot \bm\nabla \tilde\bxi = \tilde{\bu} - \barbu \quad 
    & \textrm{and} \quad \partial_t \bxi + \barbu \bm\cdot \bm\nabla \bxi = \bu \circ (\id + \bxi) - \barbu. 
\end{align}
\end{subequations}
 We solve \eqref{eq:multi_xi} and \eqref{eq:multi-scalar_PDE} to obtain the Butterworth Lagrangian mean vorticity  $\barL \zeta$ in the simulation of \S\ref{sec:shallow}.   We set $\alpha = 0.5$ and use a bilinear interpolation to evaluate $\bu$ and $g$ at the position $\bXi(\bx,t) = \bx + \bxi(\bx,t)$. Figure \ref{fig: multi-vortex_sim_HM} compares the exponential and Butterworth means by showing Hovmöller  diagrams for $\barL \xi$ at $x = 0.24$. The residual wave-induced oscillations affecting the exponential mean (top panel) are effectively eliminated by the 2nd-order Butterworth filter.  
This is confirmed in figure \eqref{fig:multi-zetafixedpoint} which shows $\barL \zeta$ for the two filters as a function of $t$ at fixed $(x,y) = (1.2,1.2)$ and by the animations \href{http://www.maths.ed.ac.uk/~vanneste/exponentialMean/movie3.mp4}{movie 3} and \href{http://www.maths.ed.ac.uk/~vanneste/exponentialMean/movie4.mp4}{movie 4} which compare the evolution of $\zeta$, the Butterworth $\barL{\zeta}$ and its Eulerian mean counterpart.

\section{Discussion}

In this paper, we propose and test efficient methods for the numerical computation of temporal Lagrangian means from simulation data. 
The methods adapt the PDE-based algorithm of KV23 to a class of time filters corresponding to convolution with a kernel $k(t)$ that consists of a sum of exponentials (truncated for $t < 0$ to ensure causality). 
These filters have the unique advantage that all the fields required in the computation satisfy evolution equations with the same time variable as that of the dynamical equations. The equations for the Lagrangian mean fields can therefore be solved on-the-fly, together with the dynamical equations, with no overheads beyond those that result from additional equations. This is in contrast with other means such as the top-hat mean used by KV23 and the more sophisticated frequency filters used by B24 which require the solution of a separate initial value problem for each of the times at which the mean fields are desired.

We use shallow-water simulation data to assess the efficacy of two filters in the class considered, namely the exponential mean and the 2nd-order Butterworth filter whose kernels involve one and two exponentials, respectively. Both filter out a large-amplitude Poincar\'e wave, but not perfectly in the case of the exponential mean fields which are polluted by residual oscillations. These are eliminated by the Butterworth filter.

The computational cost of the averaging increases rapidly with the number $N$ of exponential terms in the kernel. For a fluid in $d$-dimensions, computing the Lagrangian mean of $s$ scalar fields involves solving $N(d+s)$ time dependent PDEs. This might make kernels with $N \ge 2$  expensive. Depending on the application, the exponential or Butterworth filters may then be chosen depending on the importance attached to frequency selectivity. One limitation of these filters is their nonlinear phase response which results in distortion of the low-frequency signal. This can be mitigated -- at a cost -- by employing a larger number $N$ of exponentials.

\vspace{1\baselineskip} 
\label{SupMat}Supplementary movies are available as
\href{http://www.maths.ed.ac.uk/~vanneste/exponentialMean/movie1.mp4}{movie 1}, \href{http://www.maths.ed.ac.uk/~vanneste/exponentialMean/movie2.mp4}{movie 2}, \href{http://www.maths.ed.ac.uk/~vanneste/exponentialMean/movie3.mp4}{movie 3} and \href{http://www.maths.ed.ac.uk/~vanneste/exponentialMean/movie4.mp4}{movie 4}.
\vspace{1\baselineskip} 


We are grateful to O. B\"uhler for suggesting the use of the exponential mean for Lagrangian averaging.


\vspace{1\baselineskip} 
The authors report no conflict of interest.




\bibliography{ref}

\end{document}